\documentclass[a4paper]{article}

\usepackage{INTERSPEECH2022}
\usepackage{multirow}
\usepackage{algorithm}
\usepackage{algorithmic}
\usepackage{url}
\usepackage {xcolor}

\title{The DKU-OPPO System for the \\2022 Spoofing-Aware Speaker Verification Challenge}
\name{Xingming Wang$^{1,2}$, Xiaoyi Qin$^{1,2}$, Yikang Wang$^2$, Yunfei Xu$^3$, Ming Li$^{1,2}$\thanks{Corresponding Author: Ming Li.}}

\address{
$^1$School of Computer Science, Wuhan University, Wuhan, China \\
$^2$Data Science Research Center, Duke Kunshan University, Kunshan, China \\
$^3$Guangdong OPPO Mobile Telecommunications Corp., Ltd., Guangzhou, China
}
\email{ming.li369@dukekunshan.edu.cn}

\begin{document}

\maketitle

\begin{abstract}
  This paper describes our DKU-OPPO system for the 2022 Spoofing-Aware Speaker Verification (SASV) Challenge. First, we split the joint task into speaker verification (SV) and spoofing countermeasure (CM), these two tasks which are optimized separately. For ASV systems, four state-of-the-art methods are employed. For CM systems, we propose two methods on top of the challenge baseline to further improve the performance, namely Embedding Random Sampling Augmentation (ERSA) and One-Class Confusion Loss(OCCL). Second, we also explore whether SV embedding could help improve CM system performance. We observe a dramatic performance degradation of existing CM systems on the domain-mismatched Voxceleb2 dataset. Third, we compare different fusion strategies, including parallel score fusion and sequential cascaded systems. Compared to the 1.71\% SASV-EER baseline, our submitted cascaded system obtains a \textbf{0.21\%} SASV-EER on the challenge official evaluation set.
\end{abstract}
\noindent\textbf{Index Terms}: Anti-spoofing, Speaker verification, Spoofing Countermeasure, Spoofing-Aware Speaker Verification

\section{Introduction}

Although audio spoofing countermeasure (CM) \cite{wu2015spoofing} is highly related to Automatic Speaker Verification (ASV)\cite{bai2021speaker}, most of the research on these two tasks has been carried out independently in recent years. This may lead to CM systems not being well suited to some ASV scenarios due to overfitting or domain mismatch \cite{muller21_asvspoof}. To address this gap, the organizers of the ASVspoof Challenge \cite{todisco19_interspeech,yamagishi21_asvspoof} proposed the tandem detection cost function (t-DCF) metric \cite{kinnunen2018t}, which is highly correlated to both the ASV system and the CM system, to replace the Equal Error Rate (EER) metric, which relied only on the CM system itself. However, the ASVspoof Challenge still focuses on designing and optimizing a stand-alone CM system to calculate the min t-DCF metric in combination with a given official black-box ASV system. This prevents participants from improving the overall performance by enhancing the ASV system or leveraging joint optimization. Therefore, the first spoofing-aware speaker verification (SASV) challenge \cite{jung2022sasv}, which aims to promote the development of integrated systems that can perform both ASV and CM tasks, has been organized this year. The goal of this challenge is to build a hybrid system that can detect both zero-effort impostor access attempts and spoofing attacks simultaneously.

The 2022 SASV challenge focuses on logical access spoofing attacks (LA), such as text-to-speech (TTS) and voice conversion (VC), rather than physical access spoofing attacks (PA), such as record and play-back. Due to the lack of large scale dataset with both speaker and spoofing labels available, few studies involving joint ASV and CM optimization have been conducted in the past. Two categories of jointly optimized solutions have been summarized in the evaluation plan \cite{jung2022sasv}. The first is ensemble systems based on a fusion of separate ASV and CM systems. Gomez et al. \cite{gomez2020joint} use an embedding concatenation strategy to construct an ensemble classification system. Another approach is to build a single integrated system. Li et al. \cite{li2020joint} propose a single model using multi-task learning with contrastive loss function.

We investigated and implemented different dual-system score combination methods, including score-fusion and cascaded systems. Since the ensemble kind of solutions highly relies on the performance of the pre-trained subsystems, we investigate two innovative schemes to improve the CM system performance. The one-class confusion loss aims to reduce the intra-class Euclidean distance of bonafide audio embeddings. The random embedding sampling augmentation mechanism is also proposed for improving the model's generalization to unseen attacks. Moreover, we found that the performance of CM systems trained on ASVspoof 2019 LA \cite{todisco19_interspeech} data degraded substantially on Voxceleb2 \cite{nagrani2017voxceleb}, making the CM system more challenging to utilize the speaker embeddings trained by Voxceleb2 data. Among our explorations, the cascaded system still achieves the lowest error rate without considering the computational cost and real-time latency.

The rest of this paper is organized as follows. In section \ref{sec2}, our submitted system for the SASV challenge is represented, which mainly focuses on the CM system structure and score combination strategies. Implementation details in terms of the dataset usage and model hyperparameters are provided in Section \ref{sec3}. Section \ref{sec4} describes and discusses the results based on our experiments. Conclusions are provided in Section \ref{sec5}.

\section{System Description}
\label{sec2}
\subsection{ASV subsystem}
Four speaker verification subsystems with different network structures have been adopted in our experiments, including ResNet \cite{resnet}, SE-ResNet \cite{senet}, SimAM-ResNet \cite{simam} and ECAPA-TDNN \cite{ecapatdnn}. The global statistic pooling (GSP) is used for the ResNet subsystem while the attentive statistics pooling (ASP) \cite{asp_pooling} is used for the other three apporaches. The ArcFace\cite{arcface} which could increase intra-speaker distances while ensuring inter-speaker compactness is utilized.

\subsection{CM subsystem}
This subsection describes the basic network structure of our CM subsystems and the two proposed methods, namely one-class confusion loss and embedding random sampling augmentation.

\subsubsection{Basic network architecture}
We choose AASIST \cite{jung2021aasist} which is the challenge CM baseline as our backbone network. It contains a RawNet2 \cite{tak2021end} based encoder and an attention network based graph module. AASIST utilizes raw waveforms as the input to learn meaningful high-dimensional spectro-tempora feature maps and then extract graph nodes of feature maps in temporal and frequency domains respectively \cite{jung2021aasist}. With a stack node that learns information from all nodes, the final CM embedding is attained by concatenating various nodes' mean and maximum values. Moreover, as mentioned in \cite{tak2022automatic}, Tak et al. provide an improved architecture in which the max pooling layer of the encoded feature maps is replaced by a 2D self-attentive pooling \cite{zhu2018self}, named as AASIST-SAP.

\subsubsection{One-Class Confusion Loss function}
Although the basic models, AASIST and AASIST-SAP, obtain great results in the development and evaluation set, there is still a large performance gap since there are unseen attack algorithms in the evaluating set. Therefore, it is necessary to enhance the generalization of our system. The space of bonafide audios is relatively stable while the domain of the attack algorithms is very diverse and unpredictable. Inspired by one-class learning \cite{zhang2021one,wang2021dku}, we proposed a One-Class Confusion Loss (OCCL), which is similar to the pairwise confusion loss defined in \cite{dubey2018pairwise}.

The binary cross-entropy loss can be defined as follows:
$$\mathcal{L}_{ce}= \sum_{i} -{(y_i\log(p_i) + (1 - y_i)\log(1 - p_i))}$$
where $y_i \in \{0,1\}$ is the class label and $p_i$ is the probability output of the classifier. The anti-spoofing CM model is trained using a combined objective with the cross-entropy loss and the proposed one-class confusion loss, which is defined as:

$$\mathcal{L}_{occ}= \sum_{i} \sum_{j\neq i} ||e_i-e_j||^2$$
where $e_i$ denotes the embedding vector extracted from the bonafide audios. The purpose of this loss function is to make the Euclidean distance of all bonafide samples more compact in the embedding space. The one-class confusion loss is only applied on bonafide audios during the training process. Therefore, the final combination loss function is defined as follows:

$$\mathcal{L}= \mathcal{L}_{ce} + \lambda  \mathcal{L}_{occ}$$
where lambda is a constant hyperparameter.

\subsubsection{Embedding Random Sampling Augmentation}
Considering that the evaluation set contains many unseen logical attacks \cite{wang2020asvspoof}, we propose a fine-tuning Embedding Random Sample Augmentation (ERSA) strategy that aims to improve the robustness of the model for unknown scenarios inspired by \cite{baweja2020anomaly}. The key idea is to generate random embedding samples from the Gaussian distributions with boundary spoof embedding centers as the mean, and labeled as spoofed speech. Firstly, we initialize the embedding centers of bonafide audio and each type of spoofing audio in the development set separately based on the pre-trained model. The boundary embedding center for each type of spoofing audio is defined as the average of the bonafide embedding centers and the spoofing embedding center. During fine-tuning, the boundary embedding centers are dynamically updated based on the embeddings of current iteration. Then we randomly generated samples from $N(\hat{\mu},\Sigma)$ where $N$ is a Gaussian distribution, $\hat{\mu}$ is the boundary spoof embedding center and $\Sigma$ is the covariance matrix of spoofing embeddings calculated in advance. After each 5 epoch training, the mean and covariance matrix of embedding centers are updated.
A detailed algorithm description can be found in \cite{sasv2022dku}.

\subsubsection{Integrating Speaker Verification Embeddings}

In addition, we also explore the possibility of integrating SV embeddings into the CM model. In this case, the final CM embedding is obtained by concatenating the original CM embedding and the SV embedding together followed by two linear layers.

\subsection{System Combination}

\subsubsection{Score Fusion System}
The baseline 1 provided by the challenge organizers just generates the final SASV score by a simple score summation. However, the score distribution of the SV system and the CM system are quite different. Thus, we explore two different strategies to combine the scores, namely the normalized score multiplication approach the same way as in \cite{sasv2022dku,zhang2022new} and the score calibration and fusion approach based on the Bosaris toolkit \cite{brummer2013bosaris}.

\subsubsection{Cascaded System}

\begin{figure}[h]
  \includegraphics[width=0.5\textwidth]{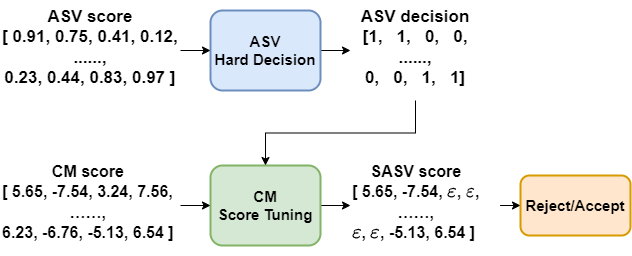}
  \centering
  \caption{{\it The illustration of the ASV followed by CM cascaded system. $\varepsilon$ represents the minimum CM score in the development set. The CM followed by ASV cascaded system is built in the same way, but switching the SV and CM systems in the pipeline.}}
  \label{cascade}
\end{figure}

As is shown in Figure \ref{cascade}, the cascaded system consists of two tandem modules: a) the first module generates a hard decision based on a threshold, the threshold is determined by the Equal Error Rate (EER) on the development set; b) the second module directly outputs the raw scores if the decision of module 1 is positive; c) the second module generates the fixed minimum score on the development set when the decision of module 1 is negative. Therefore, once the test audio is tagged as negative by the first module, the score of the second module is not useful.

For the cascaded system, two designs are considered in this work: the first one is ASV module followed by CM module, named as Cascade-ASV-CM; and the other one is CM module followed by ASV module, named as Cascade-CM-ASV. The thresholds of the first system were determined by the EER criteria on the development set.

\section{Experimental setup}
\label{sec3}

\subsection{Data Usage and Evaluation Metrics}

All datasets we used for training and validation are the training and development set of the ASVspoof2019 \cite{wang2020asvspoof} LA database and the VoxCeleb2 \cite{nagrani2017voxceleb} as requested by the organizers. The ASVspoof2019 LA database consists of bonafide and spoofing audios. Although the database contains both the speaker and spoofing labels, in general it was only used for anti-spoofing countermeasure due to the low number of  speakers. The VoxCeleb2 database contains 1128246 audios from 6112 speakers and has been widely used for ASV training. However, it is difficult to directly train a multi-speaker TTS or VC system just using Voxceleb2 \cite{Chung18b}. The official SASV evaluation trial consists of audios from the ASVspoof2019 LA evaluation partition, with unseen logical access spoofing attacks compared with audios in the train and development partitions.

The SASV-EER \cite{jung2022sasv}, which represents the EER between target and both nontarget and spoof samples, is the primary metric. SPF-EER and SV-EER are adopted as secondary metrics \cite{jung2022sasv}.

\subsection{Domain Mismatch between ASVspoof2019 LA and VoxCeleb2}
Although the AASIST-based CM system has excellent performance on the ASVspoof2019 evaluation set, it performs poorly on VoxCeleb2 based on our experiments. Most audios in VoxCeleb2 are classified as spoofing ones. We summarize two reasons that may lead to this phenomenon.

\begin{enumerate}
    \item Most audio in VoxCeleb2 contain various kinds of noises and have been coded and transmitted through some codecs and channels.
    \item The CM model trained based on the ASVspoof2019 LA dataset may learn the priori information of silent segments\cite{muller21_asvspoof}.
\end{enumerate}
The great domain mismatches between ASVspoof2019 LA and VoxCeleb2 datasets makes it difficult to improve the performance of the CM system using the VoxCeleb2 dataset. Hence, we keep those audio files in VoxCeleb2 that are classified as bonafide by our CM system as the \textbf{Vox-sub} dataset with approximately 20000 audio files in total.

\label{vox}
\subsection{Model setup}

\subsubsection{ASV subsystem}
For feature extraction, logarithmical Mel-spectrogram is extracted by applying 80 Mel filters on the spectrogram computed over Hamming windows of 20ms shifted by 10ms. The on-the-fly data augmentation \cite{cai_on-the-fly} is employed to add additive background noise or convolutional reverberation noise for the time-domain waveform. The MUSAN \cite{musan} and RIR Noise \cite{RIR} datasets are used as noise sources and room impulse response functions, respectively. We apply amplification or speed change (pitch remains untouched) to audio signals to further diversify training samples. Also, we apply speaker augmentation with speed perturbation \cite{speed_perturb_spk,dku_voxsrc20,sdsv21_qin}.  We adopt the Reduceonplateau learning rate (LR) scheduler with 0.1 initial LR. The SGD optimizer is adopted to update the model parameters.

\subsubsection{CM subsystem}

In contrast to the baseline training strategy, our trained AASIST-SAP network receives random length audio between 3-5 seconds as input. The initial learning rate is 0.001 with a Reduceonplateau learning rate scheduler. Adam optimizer is used to update the weights in models. The embedding random sample augmentation is only used during fine-tuning with two generated embeddings per center. Since there are six boundary spoof embedding centers, there will be 12 generated embeddings per batch. The batch size is set as 64 in this phase. And for the one-class confusion loss, $\lambda$ is set as 1 during training.
For fusing SV embedding into the CM model, considering the great mismatch mentioned in section 3.2, we adopt SV embeddings generated by the Resnet GSP model trained by Voxceleb2 and Voxsub as \textbf{SV-embd-V1} and \textbf{SV-embd-V2}, respectively.
\label{SVembd}
\section{Results and discussion}
\label{sec4}

\subsection{Results of the ASV Subsystem}

Table \ref{table1} reports the results of different speaker verification models. Our used models achieve state-of-the-art results on the VoxCeleb1 original test set. The ResNet with GSP achieves the best single model performance on the SASV dataset which might be because the generalization of ResNet with statistic pooling is better on this ASVspoof dataset.

\begin{table}[htb]\centering \footnotesize
    \caption{\label{table1} {\it The performances of different speaker verification subsystems on the VoxCeleb1 original test set and the SASV challenge dataset.
}}
    \begin{tabular}{cccc}
    \toprule
    \multirow{2}*{\textbf{Model}} & \multirow{2}*{\textbf{Vox-O EER[\%]}} & \multicolumn{2}{c}{\textbf{SV-EER[\%]}} \\
    \cmidrule(lr){3-4}   & &Dev &Eval \\
    \midrule
    ECAPA (Baseline) & - & 1.86 & 1.64   \\
    \midrule

    ResNet GSP & 0.851  & \bf{0.135} & \bf{0.192}   \\
    SE-ResNet34 ASP & 0.776 & 0.404 & 0.410  \\
    SimAM-ResNet34 ASP & 0.643   & 0.404 & 0.252   \\
    ECAPA-TDNN & 0.734   & 0.225 & 0.228  \\

    \bottomrule

    \end{tabular}
\end{table}

\begin{table}[htb]\centering \footnotesize
  \caption{\label{table2} {\it Comparison of different single CM systems based on SPF-EER used in SASV challenge.
  The 19LA denotes using both train and dev sets of ASVspoof2019 for training.
  The Vox-sub represents sub-bonafide audios selected from VoxCeleb2 as mentioned in Sec \ref{vox}.
  The SV-embd-V1 and V2 are defined in Sec \ref{SVembd}.
}}
  \begin{tabular}{cccc}
  \toprule
  \multirow{2}*{\textbf{Model}} & \multirow{2}*{\textbf{Data}} & \multicolumn{2}{c}{\textbf{SPF-EER[\%]}} \\
  \cmidrule(lr){3-4}   & &Dev & Eval \\
  \midrule
  AASIST(Baseline) & 19LA train & 0.07 & 0.67   \\
  \midrule
  AASIST & 19LA & 0.067 & 0.668   \\
  AASIST-SAP & 19LA  & 0.067 & 0.570   \\
  AASIST-SAP+$ERSA$  & 19LA & 0.067 & 0.510  \\
  AASIST-SAP+SV-embd-V1& 19LA  & 0.058 & 4.320  \\
  AASIST-SAP+SV-embd-V2& 19LA  & 0.067 & 1.229  \\
  AASIST-SAP & 19LA+Vox-sub   & 0.049 & 1.564 \\
  AASIST-SAP+$OCCL$ & 19LA+Vox-sub   & \bf{0.000} & \bf{0.360} \\

  \bottomrule

\end{tabular}
\end{table}

\begin{table*}[!htb]\centering \footnotesize
    \caption{\label{table3} {\it Performance of different systems evaluated in the SASV Challenge. Due to a large number of combinations, only selected combinations are listed. The $\sigma$ denotes $sigmoid$ normalization and $\times$ denotes multiplication.
  }}
    \begin{tabular}{l@{&}lllllllll}
    \toprule
    \multirow{2}*{\textbf{ID}} & \multirow{2}*{\textbf{Model}}  & \multirow{2}*{\textbf{Fusion}}  & \multicolumn{2}{c}{\textbf{SV-EER[\%]}} & \multicolumn{2}{c}{\textbf{SPF-EER[\%]}} & \multicolumn{2}{c}{\textbf{SASV-EER[\%]}} \\
    \cmidrule(lr){4-5} \cmidrule(lr){6-7} \cmidrule(lr){8-9}  & & & Dev & Eval &Dev & Eval &Dev & Eval \\
    \midrule

  CM System \\
    1 & AASIST(Baseline) & -& 46.01 & 49.24 & 0.07 & 0.67 & 15.86 & 24.38   \\
    2 & AASIST-SAP+$ERSA$ & -&47.304 & 47.188 & 0.067 & 0.510 & 15.963 & 24.655  \\
    3 & AASIST-SAP+$OCCL$ & -& 50.644 & 55.161 & \bf{0.000} & \bf{0.360} & 16.328 & 26.872   \\

    \midrule
  ASV System \\
    4 & ECAPA-TDNN (Baseline)& -& 1.86 & 1.64 & 20.28 & 30.75 & 17.31 &23.84    \\
    5 & ResNet34 GSP & -& \bf{0.135} & \bf{0.192} & 14.084 & 23.069 & 11.616 & 17.449 \\
    6 & SE-ResNet34 ASP & -& 0.404 & 0.410 & 11.540 & 22.402 & 9.745 & 16.888   \\
    7 & ECAPA-TDNN & -& 0.225 & 0.228 & 14.420 & 21.899 & 12.354 & 16.795   \\
    8 & SimAM-ResNet34 ASP & -& 0.404 & 0.252 & 12.011 & 22.500 & 10.512 & 16.994   \\

    \midrule

    & Baseline 1 (official) & Sum & 32.89 & 35.33 & 0.07 & 0.67 & 13.06 & 19.31   \\
    & Baseline 2 (official) & Back-end ensemble & 7.94 & 9.29 & 0.07 & 0.80 & 3.10 & 5.23   \\
    \midrule
    Score-fuse & ID 5+6+7 \&  ID 1+2+3 & Sum & 19.694 & 23.706 & 0.000 & 0.186 & 8.630 & 13.892 \\
     & ID 5+6+7 \&  ID 1+2+3 & $\sigma$ and $\times$ &  0.202 & 0.317 & 0.000 & 0.186 & \bf{0.103} & \bf{0.279}  \\
     & ID 5+6+7 \&  ID 1+2+3 & Bosaris & 0.134 & 0.298 & 0.009 & 0.577 & 0.067 & 0.487  \\

    \midrule

  Cascade & \multirow{2}*{ID 5 \&  ID 1 } & Cascade-ASV-CM &  0.134 & 0.204 & 0.202 & 0.477& 0.202 & 0.428   \\
          &                               & Cascade-CM-ASV &  0.202 & 0.335 & 0.067 & 0.684 & 0.149 & 0.503   \\
          \\
     & \multirow{2}*{ID 5 \&  ID 3 } & Cascade-ASV-CM & 0.135 & 0.205 & 0.135 & 0.410 & 0.135 & 0.391   \\
    & & Cascade-CM-ASV &  0.135 & 0.410 & 0.000 & 0.298 & 0.128 & 0.410   \\
\\

& \multirow{2}*{ID 8 \&  ID 2 } & Cascade-ASV-CM & 0.404 & 0.390 & 0.404 & 0.260 & 0.404 & \bf{0.288}   \\
& & Cascade-CM-ASV &  0.451 & 1.359 & 0.067 & 1.252 & 0.202 & 1.322   \\
\\

    & \multirow{2}*{ID 5+6+7  \&  ID 1+2+3} & Cascade-ASV-CM \bf{(submitted)} & 0.202 & 0.462 & 0.202 & 0.186 & 0.202 & \bf{0.209}   \\
    & & Cascade-CM-ASV & 0.173 & 0.242 & 0.000 & 0.230 & \bf{0.096} & 0.242  \\

    \bottomrule

\end{tabular}
\end{table*}

\subsection{Results of the CM Subsystem}

Table \ref{table2} shows the anti-spoofing performance of different CM subsystems. It can be seen from the table that the AASIST based model achieves a great performance improvement by replacing the max pooling layer with SAP. In addition, the model achieves a further generalizability improvement in the evaluation set by fine-tuning with ERSA strategy.

It is worth mentioning that although we extracted the Vox-sub dataset, simply adding these bonafide samples to the CM training set does not improve the CM performance. However, after integrating the proposed OCCL loss function, the overall CM performance is further enhanced. This improvement may be attributed to the fact that this loss function makes the Euclidean distances between embeddings of bonafide audios in VoxCeleb2 and ASVSpoof2019 LA closer, and thus the bonafide embedding space is more compact.

Unfortunately, simply fusing SV embedding into the CM system seems not useful. This may still be due to the domain mismatch mentioned in section 3.2, as the CM performance with SV-embd-V2 has increased considerably compared with the one with SV-embd-V1.

\subsection{Results on the Combined System}
The results of score ensemble experiments are summarized in Table \ref{table3}. More detailed results can be found in \cite{sasv2022dku}.

As can be seen from the score fusion section of the table, the simple summation method performs poorly due to the differences among the score distributions of different subsystems. This problem can be effectively mitigated by normalizing the scores through the $sigmoid$ function and multiplying them together \cite{zhang2022new}. The optimal result in this part is also obtained by this method.

The cascaded systems section of the table shows results of different cascading combinations. We have noted that the AASIST CM system in the baseline is highly complementary to the systems trained by ourselves. Furthermore, while the Cascade-CM-ASV approach performed better on the development set, the Cascade-ASV-CM approach generally performed better on the evaluation set, possibly because the development set has appeared in the training data of CM systems. In other words, for unknown scenarios, the more generalized system with lower EER is more suitable to be the first module with hard decisions. Finally, we submit the results of the Cascade-ASV-CM method.
\section{Conclusion}
\label{sec5}

In this paper, we describe our submitted system for the 2022 SASV challenge. We mainly focus on the CM subsystem and propose an embedding random sampling fine-tuning strategy to improve performance. Besides, by considering the great domain mismatch between datasets, we propose the one-class confusion loss, which improves the CM subsystem's performance even further. The final cascaded system submitted achieved 0.21\% EER on the SASV challenge evaluation set. In the future, we will try to collect a large scale database with both speaker and spoofing labels available. So we can explore more advanced joint learning approaches with sufficient data.

\section{Acknowledgements}
This research is funded in part by the National Natural Science Foundation of China (62171207), Science and Technology Program of Guangzhou City (202007030011). Many thanks for the computational resource provided by the Advanced Computing East China Sub-Center.

\bibliographystyle{IEEEtran}

\bibliography{mybib}

\end{document}